\documentclass[letterpaper,prl,twocolumn,showpacs,superscriptaddress]{revtex4}

\usepackage{graphicx}
\usepackage{times}

\begin{document}

\title{Experimental investigation of continuous variable quantum
teleportation}

\author{Warwick P. Bowen}
\affiliation{Department of Physics, Faculty of Science, Australian
National University, ACT 0200, Australia}
\author{Nicolas Treps}
\affiliation{Department of Physics, Faculty of Science, Australian
National University, ACT 0200, Australia}
\author{Ben C. Buchler}
\affiliation{Department of Physics, Faculty of Science, Australian
National University, ACT 0200, Australia}
\author{Roman Schnabel}
\affiliation{Department of Physics, Faculty of Science, Australian
National University, ACT 0200, Australia}
\author{Timothy C. Ralph}
\affiliation{Department of Physics, Centre for Quantum Computer
Technology, University of Queensland, St Lucia, QLD, 4072 Australia}
\author{Hans -A. Bachor}
\affiliation{Department of Physics, Faculty of Science, Australian
National University, ACT 0200, Australia}
\author{Thomas Symul}
\affiliation{Department of Physics, Faculty of Science, Australian
National University, ACT 0200, Australia}
\author{Ping Koy Lam}
\affiliation{Department of Physics, Faculty of Science, Australian
National University, ACT 0200, Australia}

\begin{abstract}

We report the experimental demonstration of quantum teleportation of
the quadrature amplitudes of a light field.  Our experiment was stably
locked for long periods, and was analyzed in terms of fidelity,
$\mathcal{F}$; and with signal transfer, $T_{q}=T^{+}+T^{-}$, and
noise correlation, $V_{q}=V_{{\rm in|out}}^{+} V_{{\rm in|out}}^{-}$. 
We observed an optimum fidelity of $0.64 \pm 0.02$, $T_{q}= 1.06 \pm
0.02$ and $V_{q} =0.96 \pm 0.10$.  We discuss the significance of both
$T_{q}>1$ and $V_{q}<1$ and their relation to the teleportation
no-cloning limit.

\end{abstract}


\maketitle

% \section{Introduction}

Quantum teleportation \cite{Bennett93} is a key quantum information
technology both in terms of communicating \cite{Nielsen00} and
processing \cite{Gottesman99} quantum information.  Experimental
demonstrations of teleportation have so far fallen into three main
categories: teleportation of photon states \cite{Photons}; of ensemble
properties in liquid NMR \cite{Nielsen98}; and of optical field states
\cite{Furusawa98}.  An important feature of the technique used in the
optical field state experiment of Furusawa {\it et al.}
\cite{Furusawa98} is its high efficiency.  This results in the ability
to faithfully teleport arbitrary input states continuously.  This is
due to the in principle ability to perform the required joint
measurements exactly and the technical maturity of optical field
detection.  In contrast, the efficiency of single photon experiments
is presently restricted in principle due to the inability to identify
all four Bell states, and also in practice by the low efficiency of
single photon production and detection.

Since the Furusawa {\it et al.} experiment there have been many
proposals for how quantum teleportation may be repeated using
different systems \cite{Ralph98, Leuchs99, Peng99}; applied to
different input states \cite{Polkinghorne99, Ide02}; generalized to
multi-party situations \cite{Loock00}; and more comprehensively
characterized \cite{Ralph99, Grosshans01}.  In spite of the
considerable interest, to date no new experiment has been performed
\cite{Zhang}.

This paper reports the quantum teleportation of the quadrature
amplitudes of a light beam.  Our scheme has a number of notable
differences to the previously published experiment.  The input and
output states are both analyzed by the same homodyne detector,
allowing a more consistent evaluation of their properties.  Our
experiment is based on a Nd:YAG laser that produces two squeezed beams
in two independently pumped optical parametric amplifiers (OPAs).  We
use a more compact configuration for Alice's measurements.  Finally,
the encoding and decoding of the input and output signals using a
total of 4 independent modulators.  This allows us to completely span
the phase space of the input state.

We analyze our results using the fidelity, $\mathcal{F}$, between the
input and output states, and also with signal transfer ($T_{q}$) and
noise correlation ($V_{q}$) in a manner analogous to QND analysis
\cite{Ralph98} (which we refer to as the T-V measure henceforth). 
This enables us to give a more detailed characterization of the
performance of our teleporter.

%\section{Theory}

Teleportation is usually described as the disembodied transportation
of an {\it unknown quantum state} from one place (Alice) to another
(Bob).  In our experiment, as in ref.~\cite{Furusawa98}, the
teleported states are modulation sidebands of a bright optical beam. 
The teleportation process can be described using the field
annihilation operator, $\hat a \!  = \!  (\hat X^{+} \!  + \!  i \hat
X^{-})/2 \!  $, where $\hat X^{\pm}=2\alpha^{\pm}+\delta \hat X^{\pm}$
are the amplitude (+) and phase (-) quadratures of the field,
$\alpha^{\pm} \!  = \!  \langle \hat X^{\pm} \rangle/2$ are the real
and imaginary parts of the coherent amplitudes, and $\delta \hat
X^{\pm}$ are the quadrature noise operators.  Throughout this paper
the variances of these noise operators are $V^{\pm} \!  = \!  \langle
{\delta \hat X^{\pm}}^{2} \rangle$.  The fidelity can be evaluated
from the overlap of the input (in) and output (out) states, and for
Gaussian states is given by
\begin{equation}
\mathcal{F}=2 e^{-(k^{+}+k^{-})} \sqrt{\frac{V_{{\rm in}}^{+}V_{{\rm
in}}^{-}}{(V_{{\rm in}}^{+}+V_{{\rm out}}^{+})(V_{{\rm
in}}^{-}+V_{{\rm out}}^{-})}}
\end{equation}
where $k^{\pm}\!=\!{\alpha^{\pm}_{{\rm
in}}}^{2}(1\!-\!g^{\pm})^{2}\!/(V_{{\rm in}}^{\pm}\!+\!V_{{\rm
out}}^{\pm})$, and $g^{\pm}=\alpha^{\pm}_{{\rm
out}}/\alpha^{\pm}_{{\rm in}}$ are the teleportation gains.  For a
sufficiently broad set of coherent states, the best average fidelity
at unity gain achievable without entanglement is $\mathcal{F}_{{\rm
class}}\!  =\!  0.5$.  Another interesting limit is at $\mathcal{F}
=2/3$.  This limit guarantees that Bob has the best copy of the input
state and is commonly referred to as the {\it no-cloning limit}
\cite{Grosshans01}.  Ideal teleportation would result in $\mathcal{F}
= 1$.

Alternatively, quantum teleportation can be defined as the transfer of
{\it quantum information} between Alice and Bob.  This more general
definition includes cases for which only the useful quantum features
of a system have been transferred.  In such cases a demonstrably
quantum result may be obtained even though other features of the
state, for example the state amplitude, have been distorted
sufficiently to degrade fidelity.  In the absence of entanglement,
strict limits are placed on both the accuracy of measurement and
reconstruction of an unknown state.  These are the so-called two
quantum duties.  In contrast to fidelity, they can be expressed in an
input state independent manner that is invariant under local
symplectic operations.

Bob's reconstruction is limited by the generalized uncertainty
principle of Alice's measurement $V_{M}^{+} V_{M}^{-} \!  \ge \!  1$
\cite{aut88}, where $V_{M}^{\pm}$ are the measurement penalties which
holds for simultaneous measurements of conjugate quadrature
amplitudes.  In the absence of entanglement, this places a strict limit
on Bob's reconstruction accuracy.  The limit can be expressed in terms
of quadrature signal transfer coefficients \cite{Ralph98},
$T^{\pm}={\rm SNR}^{\pm}_{{\rm out}}/{\rm SNR}^{\pm}_{{\rm in}}$ as
\begin{equation}
T_{q}=T^{+}+T^{-}-T^{+}T^{-} \left (1-\frac{1}{V_{{\rm in}}^{+}V_{{\rm
in}}^{-}} \right ) \le 1
\label{tu}
\end{equation}
where ${\rm SNR^{\pm}} \!  = \!  \alpha^{\pm}/V^{\pm}$ are the
signal-to-noise ratios. For minimum uncertainty input states ($V_{{\rm
in}}^{+}V_{{\rm in}}^{-}=1$), this expression reduces to
$T_{q}=T^{+}+T^{-}$ .

Bob's reconstruction must be carried out on an optical field, the
fluctuations of which obey the uncertainty principle.  In the absence
of entanglement, these intrinsic fluctuations remain present on any
reconstructed field.  Thus the amplitude and phase conditional
variances, $V_{{\rm in|out}}^{\pm} \!  = \!  V_{{\rm out}}^{\pm} \!  -
\!  |\langle \delta \hat X_{{\rm in}}^{\pm} \delta \hat X_{{\rm
out}}^{\pm}\rangle|^{2} \!  /V_{{\rm in}}^{\pm}$, which measure the
noise added during the teleportation process, must satisfy $V_{{\rm
in|out}}^{+} V_{{\rm in|out}}^{-} \!  \ge \!  1$.  This can be written
in terms of the quadrature variances of the input and output states
and the teleportation gain as
\begin{equation}
V_{q}=(V_{{\rm out}}^{+} - {g^{+}}^{2} V_{{\rm in}}^{+}) (V_{{\rm
out}}^{-}
- {g^{-}}^{2} V_{{\rm in}}^{-}) \ge 1
\label{vtu}
\end{equation}
It should be noted that $(V_{{\rm in|out}}^{+}\!+\!V_{{\rm
in|out}}^{-}) \ge 2$ has also been proposed for the conditional
variance limit.  For cases where both quadratures are symmetric, as
considered previously \cite{Ralph98, Polkinghorne99}, both limits are
equivalent.  The product limit, however, is significantly more immune
to asymmetry in the teleportation gain and is therefore preferred. 
The criteria of eqs.~(\ref{tu}) and (\ref{vtu}) enable teleportation
results to be represented on a T-V graph similar to those used to
characterise quantum non-demolition experiments \cite{Grangier98}. 

Both the $T_{q}$ and $V_{q}$ limits have independent physical
significance.  If Bob passes the $T_{q}$ limit, this forbids any
others parties from doing so, therefore ensuring that the transfer of
information to Bob is greater than to any other party.  This is an
`information cloning' limit that is particularly relevant in light of
recent proposals for quantum cryptography \cite{Grosshans02}. 
Furthermore, if Bob passes the $T_{q}$ limit at unity gain
($g^{\pm}\!=\!1$), then Bob has beaten the no-cloning limit and has
$\mathcal{F} \geq 2/3$.  Surpassing the $V_{q}$ limit is a necessary
pre-requisite for reconstruction of non-classical features of the
input state such as squeezing.  The T-V measure coincides with the
teleportation no-cloning limit at unity gain when both
$T_{q}\!=\!V_{q}\!=\!1$.  Clearly it is desirable that the $T_{q}$ and
$V_{q}$ limits are simultaneously exceeded.  Perfect reconstruction of
the input state would result in $T_{q} \!  = \!  2$ and $V_{q} \!  =
\!  0$.

% \section{Experiment}

\begin{figure}[!t]
    \centerline{\includegraphics[width=7.5cm]{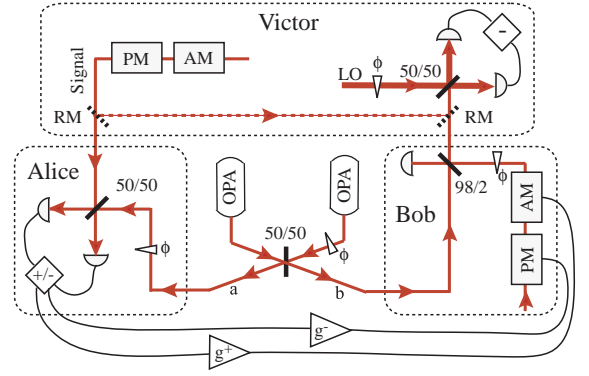}}
      \vspace{0mm} \caption{Schematic of the teleportation experiment.
      RM: removable mirror; 50/50: symmetric beam splitter; 98/2: 98\%
      transmitting beam splitter; $\phi$: phase control; A(P)M:
      amplitude (phase) modulators.}
      \label{expt}
\end{figure}
The laser source for our experiment was a 1.5~W monolithic non-planar
ring Nd:YAG laser at 1064nm.  Its output was split into two roughly
equal power beams.  One beam was mode-matched into a MgO:LiNbO$_{3}$
frequency doubler producing 370~mW of 532~nm light.  The other beam
was passed through a high finesse ring cavity to reduce spectral
noise.  This spectrally cleaned beam, which was quantum noise limited
above $6$~MHz, was then used to generate the signal for teleportation;
to seed a pair of MgO:LiNbO$_{3}$ OPAs; and to provide local
oscillator beams.

Our experimental setup to generate entanglement and perform
teleportation is shown in Fig.~\ref{expt}.  We produced the
entanglement by combining a pair of amplitude squeezed beams with a
$\pi/2$ phase shift on a 50/50 beam splitter \cite{Ou92}.  The
squeezed beams were produced by the two OPAs, each pumped with half of
the 532~nm light \cite{Bowen02}.  We characterized the entanglement
with the inseparability measure proposed by Duan {\it et al.}
\cite{Duan}; and obtained the result $ (\langle ( \hat X \!
^{+}\!_{a} \!  - \!  \hat X \!  ^{+}\!_{b})^{2} \rangle \!  + \!
\langle ( \!  \hat X \!  ^{-}_{a} \!  + \!  X \!  ^{-}_{b})^{2}
\rangle \!)/2 = \!  0.44 \pm 0.02$, where subscripts $a$ and $b$ label
the two entangled beams; this result is well below the inseparability
limit of unity.  In our situation this value becomes equivalent to the
average of the squeezed variances of the two OPAs.  This corresponds
to $3.6$~dB of squeezing on each beams.  Taking account of $16 \%$
loss in post-entanglement optics, we infer $4.8$~dB of squeezing at
the output of each OPA.

The teleportation experiment (Fig.~\ref{expt}) consisted of three
parts: measurement (Alice), reconstruction (Bob), and generation and
verification (Victor).  At the generation stage, a beam was
independently phase and amplitude modulated at 8.4~MHz.  Alice then
took one of the entangled beams and combined it on a 50/50
beamsplitter with the input state with $\pi/2$ phase shift.  The
intensities of these two beams were balanced so that the sum
(difference) of the photocurrents obtained through detection of the
two beamsplitter outputs provided a measure of the amplitude (phase)
quadrature of the input state combined with the entangled beam.  These
photocurrents were sent electronically to Bob.  Bob used them to
modulate an independent laser beam that was then combined with the
second entangled beam on a 98/2 beam splitter.  One output of this
beam splitter was Bob's reconstructed output state.

By using removable mirrors, Victor could measure the Wigner functions
of both the input and output states.  Assuming Gaussian states, Victor
need only measure the two quadratures to fully characterize the input
state.  We achieved these measurements in a locked homodyne detector. 
It is interesting to note that imperfections such as inefficiency and
low local oscillator power actually improve the results obtained by
Victor.  In our analysis, we corrected for these effects.

Active control of the entire experiment required 10 locking loops
and 4 temperature control loops.  They ensured that the mode cleaner,
frequency doubler, 2 OPA cavities, phases of the OPA pump beams, Alice
and Bob phases, Victor homodyne detection, and the relative phase
between the squeezed beams were all stably locked.
\begin{figure}[t!]
   \begin{center}
   \includegraphics[width=8cm]{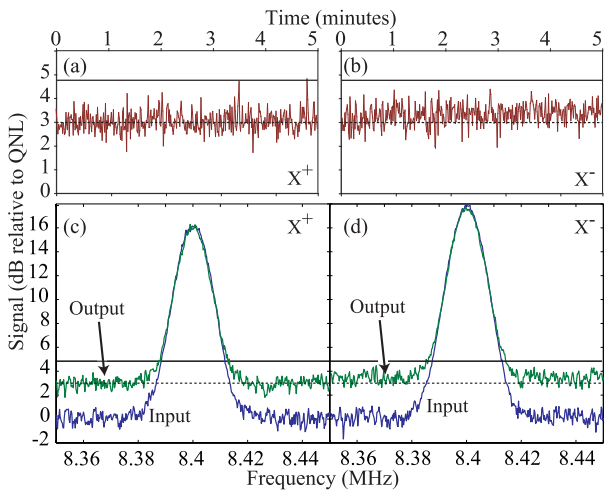}
   \end{center}
   \caption{The input and output states of the teleporter, as measured
   by Victor.  (a) and (b) show the amplitude and phase noise of the
   output state at 8.4~MHz.  (c) and (d) show the input and output of
   the teleporter, when probed with a signal at 8.4~MHz.  In all cases,
   the dotted line is the no-cloning limit, while the solid line is the
   classical limit.  All data has been corrected to account for the
   detection losses of Victor.  Resolution Bandwidth = 10~kHz, Video
   Bandwidth=30~Hz.}
   \label{in_out}
\end{figure}
A sample of the data obtained from our teleporter is shown in
Fig.~\ref{in_out}.  Parts (a) and (b) show the noise of the output
state as a function of time.  The complete system maintained lock for
long periods.  The data in (c) and (d) show the measurement of the two
quadratures over a 100~kHz bandwidth.  Over this range the noise floor
of the system was constant.  The signal-to-noise-ratio was therefore
found by comparing the peak height at 8.4~MHz to the noise at 8.35 and
8.45~MHz.  Every set of teleportation data consisted of four spectra,
such as those shown in Fig.~\ref{in_out}~(c) and (d), as well as a
quantum noise calibration (not shown).  Also drawn in each part of
Fig.~\ref{in_out} are lines corresponding to the classical limit
(solid line @ 4.8~dB) and the no-cloning limit (dashed line @ 3~dB). 
For this data set, the noise floor of both quadratures lies
convincingly below the classical limit and approaches very close to
the no-cloning limit.  Note that these limits are those calculated for
an ideal lossless teleporter.  The fidelity obtained for this data was
$\mathcal{F}\!=\!0.64 \!  \pm \!  0.02$.
\begin{figure}[!t]
    \centerline{\includegraphics[width=7.5cm]{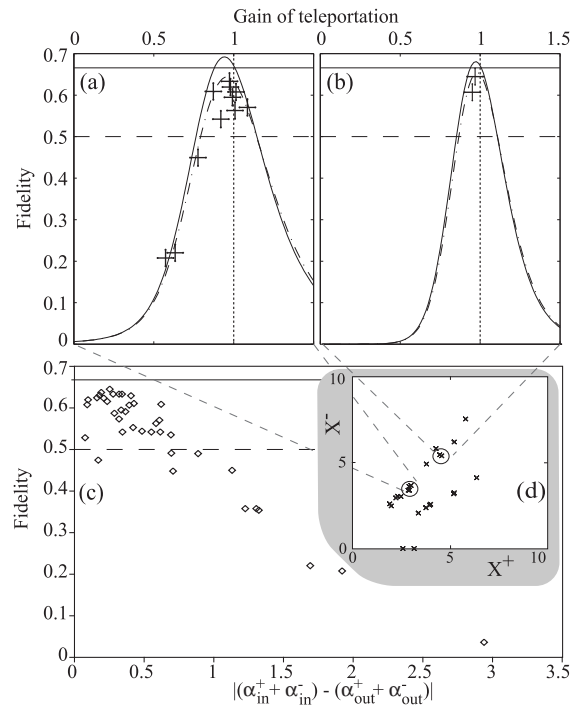}}
      \vspace{0mm} \caption{Measured fidelity plotted; versus
      teleportation gain, $g$, in (a) and (b); versus coherent
      amplitude separation between input and output states in (c); and
      on phase space in (d).  In (a) the input signal size was
      $(\alpha^{+},\alpha^{-}) \!  \approx \!  (2.9,3.5)$ and in (b)
      $(\alpha^{+},\alpha^{-}) \!  \approx \!  (4.5,5.4)$.  $g$ was
      calculated as the ratio of the input and output coherent
      amplitudes.  The dashed (solid) lines show the classical
      (no-cloning) limits of teleportation at unity gain.  The solid
      curves are calculated results based on available entanglement,
      the dot-dashed curves include the experimental asymmetric gains:
      for (a) $g^{-} \!  =0.84g^{+}$ and for (b) $g^{-}\! 
      =0.92g^{+}$.}
      \label{FvG}
\end{figure}

Fig.~\ref{FvG}(d) shows the area of phase space that our experiment
has probed.  All points shown here satisfied $\mathcal{F}\!>\!0.5$. 
For the most part, our input states had non-zero coherent amplitude
components, thereby allowing verification of the gains of both
quadratures.  One of the features of fidelity is a strong dependence
on gain and signal size.  For example, in the limit of a vacuum input
state, the fidelity criterion will actually be satisfied perfectly by
a classical teleporter (i.e. one with the entangled state replaced by
two coherent states) with zero gain.  The fidelity criterion therefore
requires proof that the gain of a teleportation event is unity.  A
subset of our data is shown in Fig.~\ref{FvG} (a) and (b).  Each
diagram plots fidelity as a function of teleportation gain for results
with identical input states.  The solid curves show the best possible
performance of our system, based on our entanglement, detection
efficiency, dark noise, and assuming equal gain on each quadrature. 
Both plots demonstrate that the highest fidelity occurs for gain less
than unity.  The increased fidelity is less obvious in (b) where the
signal is approximately twice as large as that in (a).  For small
signals it is therefore crucial to ensure unity gain.  Obtaining the
correct gain setting is actually one of the more troublesome
experimental details.  To illustrate this point, we have plotted the
dashed curves on (a) and (b) for a teleporter with asymmetric
quadrature gains.  Such asymmetry was not unusual in our system, and
explains the variability of the results shown in Fig.~\ref{FvG}(a).  A
summary of all our fidelity results is shown in Fig.~\ref{FvG}(c) as a
function of deviation from unity gain.

\begin{figure}[!t]
    \centerline{\includegraphics[width=7.5cm]{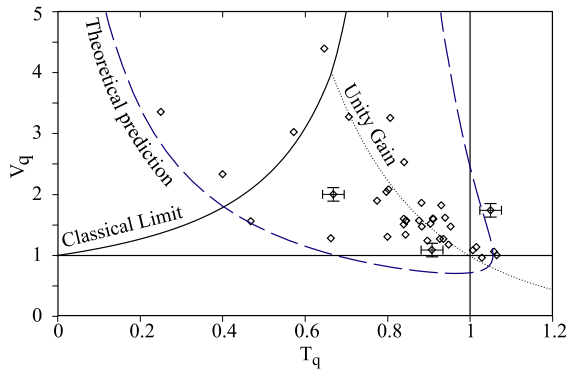}}
      \vspace{0mm} \caption{T-V graph of the experimental results. 
      The dashed theoretical curve was calculated based on the
      available entanglement and experimental losses.  Representative
      error bars are shown for some points.}
      \label{T-V}
\end{figure}

Analyzing teleportation results on a T-V graph has several advantages
when compared to fidelity.  The T-V graph is two dimensional, and
therefore conveys more information about the teleportation process.
It tracks the quantum correlation and signal transfer in non-unity
gain situations.  In particular, it identifies two particularly
interesting regimes that are not evident when using fidelity: the
situations where the output state has minimum noise (minimum $V_{q}$),
and when the input signals were transferred to the output state
optimally (maximum $T_{q}$).  Our T-V results are shown in
Fig.~\ref{T-V}.  The classical limit curve shows the ideal achievable
result as a function of gain if the entanglement was replaced with two
coherent states.  The unity gain curve shows the locus of points
obtained at unity teleportation gain with increasing entanglement.
Finally, a theoretical curve (as a function of gain) is shown for our
experimental parameters.  By varying our experimental conditions,
particularly the gain, we have mapped out some portion of the T-V
graph.  

Perhaps the most striking feature of these results are the points with
$T_q>1$, the best of which has $T_q=1.06\pm0.03$.  Since only one
party may have $T_q>1$, this shows that Bob has maximal information
about the input signal and we have broken the information cloning
limit.  The lowest observed conditional variance product was
$V_{q}\!=\!0.96 \!  \pm \!  0.10$.  This point also had $T_{q}=1.04 \! 
\pm \!  0.03$.  This is the first observation of both $T_{q}>1$ and
$V_{q}<1$, and with unity gain this would imply breaking of the
no-cloning limit for teleportation.  This particular point, however,
had a fidelity of $0.63 \pm 0.03$.  The main reason for this low
fidelity is asymmetric gain, the amplitude gain was $ g^{+}\!  = \! 
0.92 \pm 0.08$ while the phase gain was $g^{-} \!  = \!  1.12 \pm
0.08$.  Such gain errors have a dramatic impact on the measured
fidelity because the output state then has a different classical
amplitude ($\alpha^{\pm}$) to the input, a difference in the classical
properties of the input and output states to which fidelity is very
sensitive.

In conclusion, we have performed stably locked quantum teleportation
of an optical field.  The best fidelity we directly observed was
$\mathcal{F}\!=\!0.64 \!  \pm \!  0.02$.  The maximum two quadrature
signal transfer for our apparatus was $T_{q}\!=\!  1.06 \!  \pm \! 
0.03$.  We also observed a conditional variance product of
$V_{q}\!=\!0.96 \!\pm\!  0.10$ coincident with $T_{q}\!=\!  1.04 \! 
\pm \!  0.03$.  This is the first observation of both $T_{q}>1$ and
$V_{q}<1$.  At unity gain this would ensure violation of the
no-cloning limit for teleportation.  The asymmetry in our gain,
however, prevented a direct measurement of $\mathcal{F}>0.67$, thereby
leaving the no-cloning limit as a tantalising prospect for future
experiments.

We thank the Australian Research Council for financial support and the
Alexander von Humboldt foundation for support of R.~Schnabel.  This
work is a part of EU QIPC Project, No.~IST-1999-13071 (QUICOV).

\end{document}